\documentclass{pnastwo}

\usepackage[dvips]{graphicx}

\usepackage{amssymb,amsfonts,amsmath}

\begin{document}



\conflictofinterest{The authors declare no conflict of interest.}

\track{This paper was submitted directly to the PNAS office.}


\footcomment{Abbreviations: FL, Fermi liquid; n-FL, non-Fermi liquid; QCP, quantum critical point }


\title{Field-induced quantum critical route to a Fermi liquid in
high-temperature superconductors}




\author{
Takasada Shibauchi\affil{1}{Department of Physics, Kyoto University, Sakyo-ku, Kyoto 606-8502, Japan}\thanks{To whom correspondence should be addressed. E-mail:shibauchi@scphys.kyoto-u.ac.jp},
Lia Krusin-Elbaum\affil{2}{IBM T.J. Watson Research Center, Yorktown Heights, New York 10598},
Masashi Hasegawa\affil{3}{Department of Materials Science and Engineering, Nagoya University, Chikusa-ku, Nagoya 464-8603, Japan},
Yuichi Kasahara\affil{1}{},
Ryuji Okazaki\affil{1}{}, \and
Yuji Matsuda\affil{1}{}\affil{4}{Institute for Solid State Physics, University of Tokyo, Kashiwa, Chiba 277-8581, Japan}
}

\contributor{Submitted to Proceedings of the National Academy of Sciences
of the United States of America}

\maketitle

\begin{article}

\begin{abstract}
In high transition temperature ($T_{\rm c}$) superconductivity,
charge doping is a natural tuning parameter that takes copper oxides
from the antiferromagnet to the superconducting region.
In the metallic state above $T_{\rm c}$ the standard Landau's
Fermi-liquid theory of metals as typified by the temperature squared
($T^2$) dependence of resistivity appears to break down. Whether the
origin of the non-Fermi-liquid behavior is related to physics
specific to the cuprates is a fundamental question still under
debate. We uncover a new transformation from the non-Fermi- to a
standard Fermi-liquid state driven not by doping but by magnetic
field in the overdoped high-$T_{\rm c}$ superconductor
Tl$_2$Ba$_2$CuO$_{6+x}$. From the $c$-axis resistivity measured up
to 45~T, we show that the Fermi-liquid features appear above a
sufficiently high field which decreases linearly with temperature
and lands at a quantum critical point
near the superconductivity's upper critical field ---
with the Fermi-liquid coefficient of the $T^2$ dependence showing a
power-law diverging behavior on the approach to the critical point.
This field-induced quantum criticality bears a striking resemblance
to that in quasi-two dimensional heavy-Fermion superconductors,
suggesting a common underlying spin-related physics in these
superconductors with strong electron correlations.
\end{abstract}




\dropcap{Q}uantum criticality refers to a phase transition process
between competing states of matter governed not by thermal but by
quantum fluctuations demanded by Heisenberg uncertainty principle
\cite{Sachdev}. It has emerged at the front and center of the
physics of strongly correlated electron systems known to host
competing quantum orders, and is witnessed by a proliferation of
reports on heavy Fermions \cite{Sidorov,Nakajima,Mathur,Gegenwart},
itinerant (quantum) magnets \cite{Sr3Ru2O7}, and
high-transition-temperature (high-$T_{\rm c}$) superconductors
\cite{Valla}, with quantum matter tuned (at times arguably) through
a transition by pressure, magnetic field, or doping. Arguably, since
one has to rely on long shadows cast by quantum criticality far above zero
temperature \cite{Sudip1}, for, obviously, $T = 0$~K cannot ever be
attained.

The often-invoked hallmark of quantum criticality is an
unconventional behavior of resistivity. For resistivity
contribution, the standard Fermi liquid (FL) theory of metals
predicts a quadratic temperature dependence $\rho(T) = \rho(0)+AT^2$
at low temperatures. In high-$T_{\rm c}$ cuprates, however, the
baffling $T$-linear resistivity over a huge temperature range near
optimal (hole) doping has been observed \cite{Kubo}, flagging, in
this sense, a non-Fermi liquid (n-FL) behavior in the metallic state
above $T_c$. This has led to new theoretical concepts, some related
(e.g., phenomenology of ``marginal Fermi liquid" \cite{Varma}) and
some unrelated (e.g., ``strange metal" state \cite{Anderson2006}) to
quantum criticality. In most considerations of cuprates near quantum
critical points (QCPs) the tuning parameter is charge doping
\cite{Sachdev,Varma}. And while there is some experimental support
\cite{Valla,Tallon} for a doping-driven QCP, it is still to be
broadly confirmed. Thus, it is of primary import to probe
experimentally how the n-FL state transforms into the conventional
FL state, and whether and how charge or spin degrees of freedom are
involved.

Here we report on the transformation from such `strange' n-FL state
to the conventional FL metallic state in high-$T_{\rm c}$
superconductors in high magnetic fields. Our experiments measuring
charge transport in overdoped Tl$_2$Ba$_2$CuO$_{6+x}$ reveal an
unanticipated quantum criticality in a cuprate that is not doping-
but field-induced. The results are in close correspondence with the
quantum criticality in the quasi-two dimensional heavy-Fermion
superconductors having strong antiferromagnetic fluctuations,
suggesting common fundamental physics of magnetic origin responsible
for the observed QCP.

To have access to large regions of the metallic regime at low
temperatures, we use magnetic fields to destroy superconductivity in
heavily doped Tl$_2$Ba$_2$CuO$_{6+x}$ ($T_{\rm c}\approx 15$~K).
This material has a single CuO$_2$ layer per unit cell
and is relatively clean among cuprates as
evidenced by the high $T_{\rm c}$ (up to 93~K) that can be
controlled with oxygen content. We focus here on the $c$-axis
longitudinal magnetotransport ($H \parallel c$), since it should be
less affected by orbital contributions than the transverse geometry,
and since in our overdoped system it is expected that Fermi surface
is three-dimensional-like and coherent \cite{Hussey}, as revealed by
the fact that the temperature dependence of $c$-axis resistivity
$\rho_c(T)$ can be well scaled by that of $ab$-plane resistivity
$\rho_{ab}(T)$ [see below].

\section{Results and Discussion}

Figure~\ref{rho_T} shows the temperature dependence of $c$-axis
resistivity at zero and 45~T fields. At zero field, $\rho_c(T)$ is
metallic all the way down to $T_{\rm c}$. This represents a clear
contrast with the semiconductinglike upturn in $\rho_c$ observed at
lower dopings of Bi$_2$Sr$_2$CaCu$_2$O$_{8+y}$
\cite{Shibauchi2001,Krusin} in the pseudogap state \cite{Basov}. We
can examine our data within the overall temperature dependence
$\rho_c(T) = \rho_{c0}+A_0T^2+CT$ which reproduces the temperature
dependence of $\rho_{ab}(T)$ \cite{Abdel}. Also, it can be as easily
fitted by a power law with the exponent 1.3 ($<2$) (inset in
Fig.~\ref{rho_T}).
Regardless of the choice, the temperature dependence
is not $T$-quadratic as in a conventional FL;
it marks an n-FL state even in the heavily overdoped region.

When we apply 45~T along the $c$ axis, the superconductivity is
destroyed and the entire temperature dependence up to 100~K can now
be fitted with the simple FL form $\rho_{c}(0)+AT^2$. This clearly
demonstrates that sufficiently high magnetic fields destroy all
remnants of the n-FL behavior, recovering the all familiar
Fermi-liquid metal; {\it i.e.} in this overdoped cuprate there
exists \emph{a field-induced transformation from the n-FL to FL
state}.

\begin{figure}
\centerline{\includegraphics[width=90mm]{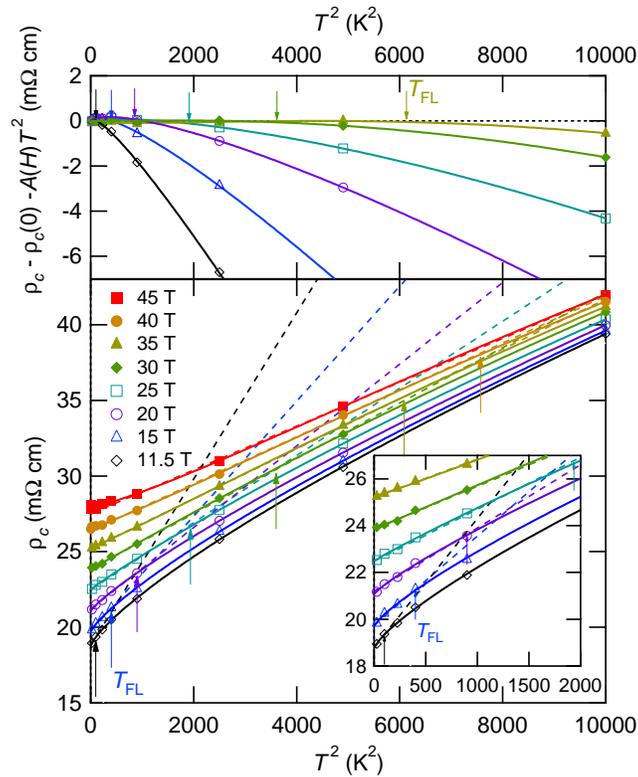}}%
\caption{$c$-axis resistivity $\rho_c$ as a function of $T^2$ at fixed fields.
Upper panel: $\rho_c$ with the Fermi-liquid contribution subtracted highlights
the non-Fermi-liquid behavior for $T>T_{\rm FL}$ (marked by
arrows). Lower panel: $\rho_c$ fitted to the $AT^2$ dependence
(dashed lines) for $T<T_{\rm FL}$.
Onsets of the deviation from $AT^2$
have error bars indicated in Fig.~\ref{T_H}. Inset shows an expanded
view of the low temperature region.} \label{rho_T2}
\end{figure}

\begin{figure}
\centerline{\includegraphics[width=85mm]{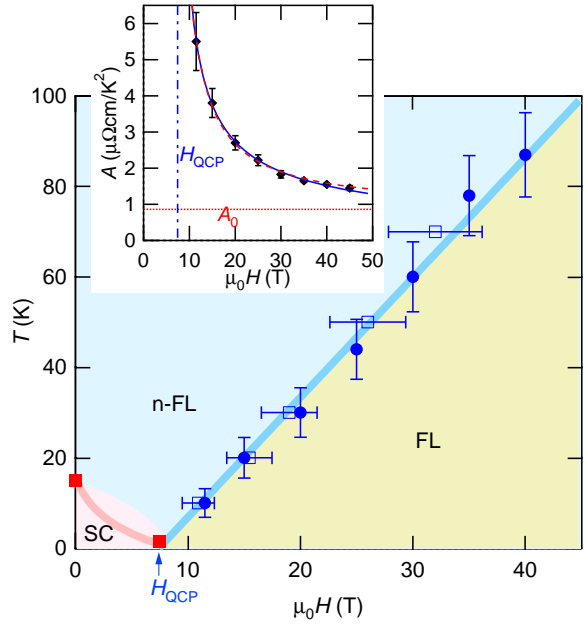}}
\caption{Temperature-field diagram obtained from the
high-field transport measurements. Blue solid circles,
$T_{\rm FL}(H)$, and open squares, $H_{\rm FL}(T)$,
separate Fermi-liquid (FL) and non-Fermi-liquid (n-FL) states.
Red squares are the onset of
superconductivity (SC). Thick red line represents $H_{\rm sc}(T)$,
which in cuprates varies exponentially with $T$
\cite{Shibauchi2001,Morozov}. Red hatched area outlines $H_{\rm
c2}(T)$. Inset: The Fermi-liquid coefficient $A$ as a function of
$H$. The data can be fitted to Eq.~(1) as shown by blue-solid and
red-dashed lines corresponding to two choices of $A_0$, see text.
} \label{T_H}
\end{figure}

To follow the temperature dependence of $\rho_c$ at different fields
we plot it against $T^2$ in Fig.~\ref{rho_T2}. It is evident that
the $AT^2$ dependence is observed below a field-dependent
temperature $T_{\rm FL}$ indicated by the arrows. At higher
temperatures the $\rho_c(T)$ data deviate from the $T^2$ behavior as
can be seen more clearly by subtracting $\rho_c(0)+A(H)T^2$ in the
upper panel. We note that although the change is gradual, the power
in the temperature dependence unmistakably changes from 2 at low
temperatures ($T<T_{\rm FL}$) to less than 2 at high temperatures
($T>T_{\rm FL}$). The field dependence of the $T_{\rm FL}$ is
depicted in the $T$-$H$ diagram in Fig.~\ref{T_H}. At 45~T, FL state
extends up to 100~K, and at lower fields the Fermi liquid breaks
down crossing to an n-FL behavior above $T_{\rm FL}$. With
decreasing field $T_{\rm FL}(H)$ decreases linearly and extrapolates
to zero in the vicinity of the upper critical field $H_{\rm c2}(0)$
[see below], terminating at a putative QCP. We conclude then that in
zero temperature limit, the normal state above $H_{\rm c2}$ in
Tl$_2$Ba$_2$CuO$_{6+x}$ is a Fermi liquid, in agreement with the
recent observation in this system of the Wiedemann-Franz law
\cite{Proust}.

Next we examine the field dependence of $\rho_c$ at constant
temperatures, plotted in Fig.~\ref{rho_H}(a). At low temperatures,
the resistivity is zero below the so-called irreversibility field
$H_{\rm irr}$ \cite{Shibauchi2001} in the vortex solid state. We
note that above $H_{\rm irr}$ the magnetoresistance is always
\emph{positive}. We recall that in less-doped pseudogapped
Bi$_2$Sr$_2$CaCu$_2$O$_{8+y}$ the observed magnetoresistance is
\emph{negative} over a large field range \cite{Shibauchi2001},
consistent with filling of the low-energy states within the
pseudogap in the applied magnetic field. We surmise then, that at
this doping the pseudogap is either way below the superconducting
energy scale, or perhaps entirely absent.

The superconducting coherence can survive up to a characteristic
field $H_{\rm sc}$, above which the quasiparticle conductivity
overcomes the vortex contribution
\cite{Morozov,Shibauchi2003,Mackenzie}. This often underestimates
the upper critical field $H_{\rm c2}$ near $T_{\rm c}$ in
high-$T_{\rm c}$ cuprates; it is notoriously difficult to obtain
from transport owing to large thermal fluctuations.
However, previous studies of $c$-axis
magnetotransport \cite{Shibauchi2003} revealed that in the overdoped
regime in the low-$T$ limit, $H_{\rm c2}$ is very near $H_{\rm
sc}(0)$. In our sample we evaluate $\mu_0H_{\rm c2}(0)\sim 8$~T.
Above this limiting field $\rho_c(H)$ at low-$T$ is strictly
$H$-linear in the normal state over the entire field range.

To take a closer look at higher $T$, we subtract the high-field
linear term from $\rho_c(H)$ and obtain $\delta \rho_c$, which
quantifies the deviation from the $H$-linear dependence. This
analysis highlights a noticeable deviation from the field-linearity
below a temperature-dependent characteristic field $H_{\rm FL}$, see
Fig.~\ref{rho_H}(b). The obtained $H_{\rm FL}(T)$ is also plotted in
Fig.~\ref{T_H}. Remarkably and \emph{consistently} it follows the
$T_{\rm FL}(H)$ line within the experimental error bars. We surmise
then, while the $H$-linear and large magnetoresistance is a
non-trivial finding in its own right that needs to be further
understood, here it clearly is a phenomenon of the Fermi-liquid.
Indeed, several theoretical accounts within the Fermi-liquid picture
derive large $H$-linear $\rho_c(H)$ \cite{Abrikosov,Schofield}.

We remark that at low temperatures below 5~K the standard FL state
is confirmed by the classical Kohler's rule for magnetoresistance,
see Fig.~\ref{rho_H}(c). At higher temperatures, where the low field
data below $H_{\rm FL}$ [including $\rho_c^{\rm n}(0)$] no longer
follow what is expected in the simple FL state, the scaling is
clearly violated. And while the violation of Kohler's rule at high
temperatures can be caused by other mechanisms, the low temperature
data are consistently in correspondence with the field-induced FL
state. The temperature-dependent violation further indicates that
here the magnetoresistance is not simply governed by $\omega_c\tau$
(a product of the cyclotron frequency and scattering time). From
this we conclude that the observed field-induced $AT^2$ behavior is
an intrinsic effect and not an artifact due to $\omega_c\tau$.

At finite temperatures the observed field-induced transformation
appears to be crossover-like. So now we will ask whether the $T
\rightarrow 0$~K terminus of $H_{\rm FL}(T)$ indicates a true phase
transition at QCP. We note the conspicuously strong field dependence
of the FL coefficient $A$: it increases with decreasing field and
decreasing $T_{\rm FL}$, see inset of Fig.~\ref{T_H}. Indeed we find
that the field dependence can be fitted to
\begin{equation}
A(H)=A_0+ D(H-H_{\rm QCP})^{-\alpha},
\end{equation}
where $A_0$ and $D$ are constants and $\alpha$ and $H_{\rm QCP}$ are
the relevant parameters of the fit. As we discussed earlier, in zero
field $\rho_c(T)$ can be analyzed either by a power law or by the
$\rho_{c0}+A_0T^2+CT$ dependence. In the analysis of the field
dependence the two different forms would require different values of
$A_0$ in Eq.~(1). In the former case, we take
$A_0=0$~$\mu\Omega$cm/K$^2$ and we can fit the $A(H)$ by $\alpha=0.62$
and $H_{\rm QCP}=7.4$~T. In the
latter case, we use the
finite coefficient $A_0=0.86$~$\mu\Omega$cm/K$^2$ (see Fig.~\ref{rho_T})
and the fit gives $\alpha=1.04$ and $H_{\rm QCP}=5.8$~T.

Thus, experimental $A(H)$ algebraically diverges at $H_{\rm QCP}$.
Within Fermi liquid theory, as $T \rightarrow 0$~K the energy
dependence of the total scattering rate near the Fermi surface takes
the form $1/\tau = 1/\tau_0 + a (E-E_{\rm F})^2$ ($1/\tau_0$ comes
from the impurity scattering, $a$ is a constant in energy $E$, and
$E_{\rm F}$ is the Fermi energy) \cite{Ashcroft}. At the QCP the
singularity of $a(H)$ will mirror that of $A(H)$ --- the two
coefficients are related through the quasiparticle-quasiparticle
scattering cross section. The found divergence of $A$ thus gives us
confidence in assigning $H_{\rm QCP}$ as the QCP field, and $\alpha$
as the exponent characterizing quantum criticality. We remark that
strongly correlated electron systems commonly obey Kadowaki-Woods
relation $A \propto \gamma^2$ \cite{Kadowaki}, where $\gamma$ is the
electronic coefficient of specific heat and a measure of the
effective mass $m^*$ of a Landau quasiparticle. While this relation
is complex (and sometimes violated \cite{Gegenwart}), we note that
with large ($\sim~10^3$ \cite{Abdel}) resistivity anisotropy in
Tl$_2$Ba$_2$CuO$_{6+x}$, the obtained $A$ values near $H_{\rm QCP}$
imply enhanced $\gamma \sim 30$~mJ/mol-K$^2$, comparable to that
e.g. in superconducting Sr$_2$RuO$_4$ \cite{Maeno},
where similarly anisotropic $A$ values between the $c$-axis and
in-plane resistivities have been observed. This enhancement
of $A$ and a lack of saturation may also be related to the enhanced
susceptibility $\chi_0$ in the overdoped
Tl$_2$Ba$_2$CuO$_{6+x}$ \cite{Kubo}. 
We surmise then that at finite temperatures the system is governed
by the quantum fluctuations, generating the n-FL state which crosses
over to the conventional FL above $H_{\rm FL}$.


The n-FL state with non-$T^2$ dependence of resistivity
\cite{Sidorov} and a violation of Kohler's rule \cite{Nakajima} has
also been observed in heavy-Fermion superconductors having strong
antiferromagnetic fluctuations. Notably, in CeCoIn$_5$ with
quasi-two dimensional electronic structure, a quite similar
field-induced QCP has been identified by the transport and specific
heat measurements \cite{Paglione,Bianchi,Movshovich}. In high
magnetic fields, the resistivity recovers the $AT^2$ dependence at
low temperatures in a similar manner near the upper critical field
$H_{\rm c2}(0)$ ($\approx 5$~T). It has been pointed out
\cite{Bianchi} that the underlying antiferromagnetic fluctuations
\cite{Moriya} become critical in the immediate vicinity of the
superconductivity, preventing development of magnetic order. We note
that anisotropic violation of the Wiedemann-Franz law in CeCoIn$_5$
was recently found near the QCP \cite{Tanatar}, where the FL
renormalization parameter $Z$ $(\sim 1/m^*)$ tends to zero in the
$c$ direction but remains finite in the $ab$ plane. 
This suggests that the $c$ direction is more susceptible to
instabilities related to QCP.

An intriguing question to ask is whether field-induced $H_{\rm QCP}
\approx H_{\rm c2}(0)$ in a highly overdoped cuprate is a shear
coincidence or are they inherently linked. In particular, one may
ask whether an extended regime of superconducting fluctuations can
promote the observed n-FL state. In the heavy-Fermion superconductor
CeCoIn$_5$, FL coefficient $A$ also diverges at the QCP located very
near $H_{\rm c2}(0)$, with $\alpha$ close to unity \cite{Paglione}.
By applying pressure $H_{\rm QCP}$ is strongly suppressed and is no
longer coincident with $H_{\rm c2}(0)$ \cite{Movshovich}. This
rather compellingly points to a QCP controlled by a competing order,
most likely related to antiferromagnetism \cite{Pham}. In cuprates,
neutron scattering experiments \cite{Lake,Wilson} show that magnetic
field can \emph{induce} a distinct static magnetic order, and a
surprisingly much enhanced spin fluctuations at low $T$ within the
vortex cores, also detected by a spatially resolved NMR
\cite{vortexNMR}. Thus, spin correlations in cuprates seem to
experience a field-induced boost.

Our work, in a departure from previous studies, probes the
high-field regime at very high hole doping---much distanced from the
antiferromagnetic `mother order'. That the antiferromagnetic
fluctuations \cite{Scalapino} could have such long reach
\cite{Wakimoto} and play a role in the uncovered field-induced QCP
is quite extraordinary. We expect that the true nature of the
quantum critical fluctuations that produce the n-FL state in the
highly overdoped Tl$_2$Ba$_2$CuO$_{6+x}$ is complex, since here we
are not far from the superconductivity's charge-doping end point
\cite{Kopp}. From our experiments, with salient similarities found
between a cuprate and a heavy-Fermion compound, all evidence here
points to a spin-controlled QCP universal to these strongly
correlated electron systems.

\section{Materials and Methods}
\begin{figure}
\centerline{\includegraphics[width=90mm]{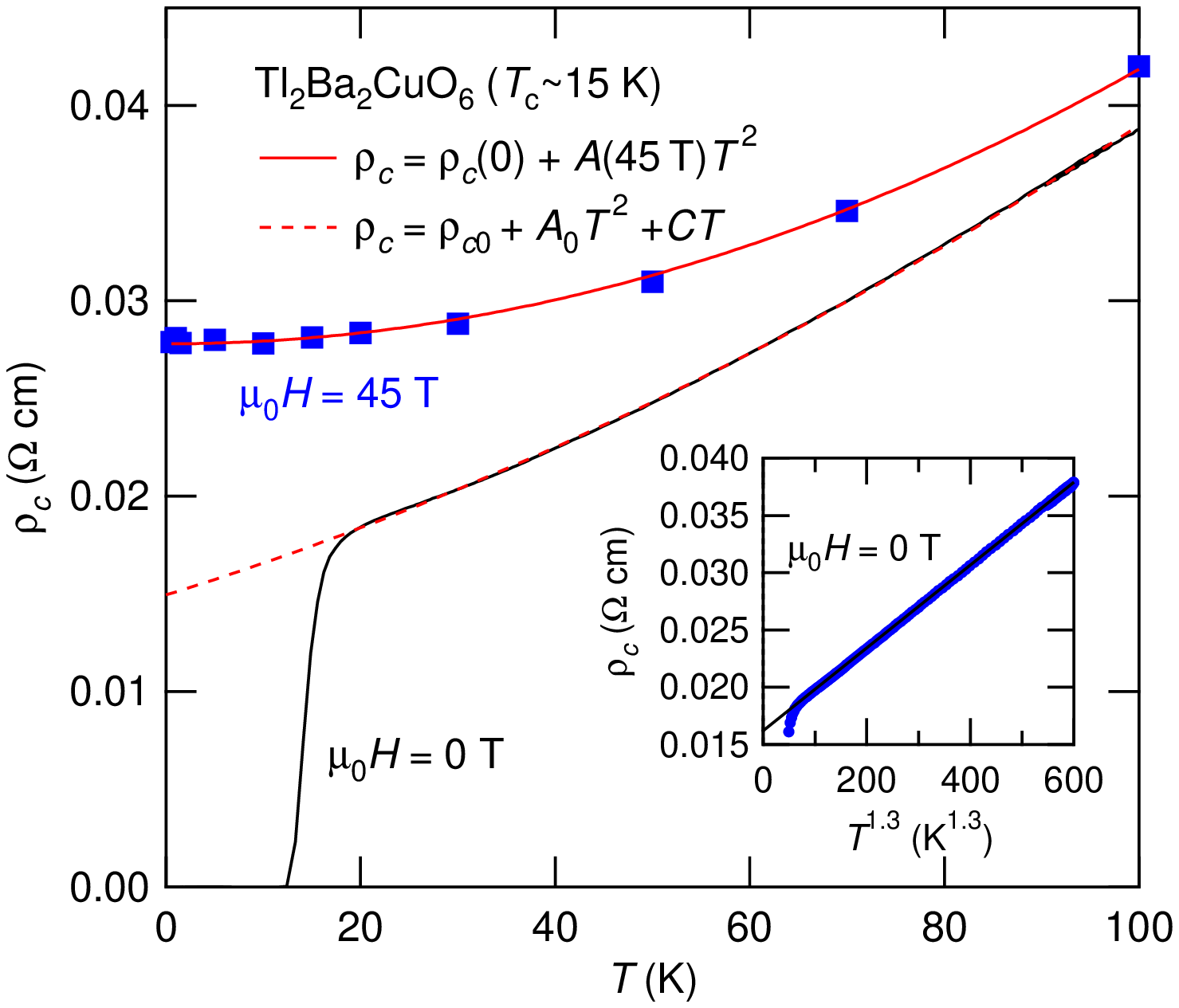}}%
\caption{Temperature dependence of the $c$-axis
resistivity $\rho_c$ in an overdoped crystal of
Tl$_2$Ba$_2$CuO$_{6+x}$ under zero (black solid line) and a 45~T
field (squares). Red dashed and solid curves are the fits to
$\rho_{c0}+A_0T^2 +CT$ and $\rho_c(0)+AT^2$, respectively. Inset:
$\rho_c$ vs $T^{1.3}$ at zero field. Solid line is a linear fit. }
\label{rho_T}
\end{figure}

\begin{figure}
\centerline{\includegraphics[width=93mm]{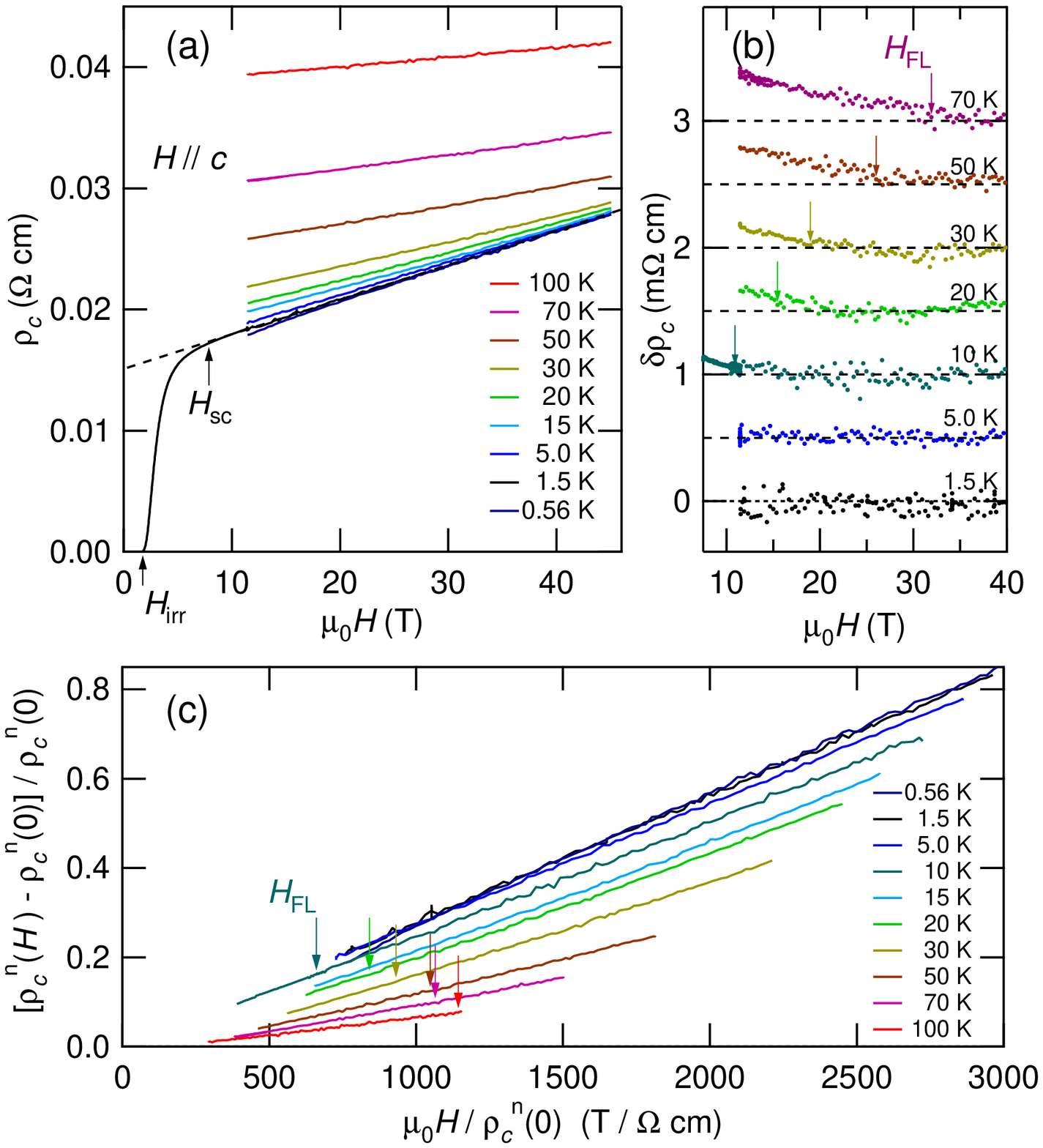}}%
\caption{Field dependence of the $c$-axis resistivity $\rho_c$.
(a) $\rho_c$ vs field $H$
at fixed temperatures. Dashed line is a linear fit to the 1.5 K
data. Below $H_{\rm sc}$ the downward rounding of
$\rho_c$ signifies the onset of superconductivity,
and $\rho_c$ is zero below the irreversibility field $H_{\rm irr}$.
(b) $\delta\rho_c(H)$ obtained by subtracting the $H$-linear part
from $\rho_c(H)$ at fixed $T$. Each curve is shifted vertically for
clarity. $H_{\rm FL}(T)$, marked by arrows [also in (c)], are the
deviation points from $H$-linear magnetoresistance (MR). (c) Kohler
plot of normal-state MR
against $\mu_0H/\rho_c^{\rm n}(0) \propto \omega_c \tau$.
$\rho_c^{\rm n}(0)$ is the normal-state zero-field $\rho_c(T)$
[dashed line in Fig.~\ref{rho_T}].
}
\label{rho_H}
\end{figure}

Single crystals of Tl$_2$Ba$_2$CuO$_{6+x}$ were grown by a flux
method \cite{Hasegawa}. In this system, the doping can be tuned by
oxygen content covering a range from somewhat overdoped ($T_{\rm
c}\approx 93$~K) up to heavily overdoped ($T_{\rm c}\approx 0$~K)
\cite{Kubo}. In our study, we used a homogeneous highly-overdoped
crystal with a sharp transition at $T_{\rm c}\approx 15$~K (see
Fig.~\ref{rho_T}). The $c$-axis resistivity $\rho_c(T,H)$ was
measured in the 45-T hybrid magnet at NHMFL (comprising a 11.5~T
superconducting outsert and 33.5~T resistive insert magnets)
by the standard four-probe method using an ac resistance bridge
\cite{Kawakami}. The temperature at high fields
was controlled to $\sim 50$~mK by using a capacitance censor
at low temperatures, where the magnetoresistance
of Cernox resistive sensors is not negligible.

\begin{acknowledgments}
{\bf ACKNOWLEDGMENTS.} We thank A.~I. Buzdin, S. Chakravarty, S.
Fujimoto, N.~E. Hussey, H. Kontani, and C.~M. Varma for discussions,
and B. Brandt for technical assistance at NHMFL. This work was
supported in part by Grants-in-Aid for Scientific Research from
JSPS, and for the 21st Century COE ``Center for Diversity and
Universality in Physics" from MEXT, Japan.
\end{acknowledgments}



\end{article}

\begin{thebibliography}{}
\bibitem{Sachdev}
Sachdev S (1999) in
{\it Quantum Phase Transitions} (Cambridge Univ. Press, Cambridge).

\bibitem{Sidorov}
Sidorov VA {\it et al.} (2002)
Superconductivity and quantum criticality in CeCoIn$_5$.
{\it Phys Rev Lett} 89:157004.

\bibitem{Nakajima}
Nakajima Y {\it et al.} (2007)
Non-Fermi liquid behavior in the magnetotransport of Ce$M$In5 ($M$: Co and Rh): Striking similarity between quasi two-dimensional heavy Fermion and high-$T_{\rm c}$ cuprates.
{\it J Phys Soc Jpn} 76:024703.

\bibitem{Mathur}
Mathur ND {\it et al.} (1998)
Magnetically mediated superconductivity in heavy Fermion compounds.
{\it Nature} 394:39-43.

\bibitem{Gegenwart}
Custers J {\it et al.} (2003)
The break-up of heavy electrons at a quantum critical point.
{\it Nature} 424:524-527.

\bibitem{Sr3Ru2O7}
Grigera SA {\it et al.} (2004)
Disorder-sensitive phase formation linked to metamagnetic quantum criticality.
{\it Science} 306:1154-1157. %

\bibitem{Valla}
Valla T {\it et al.} (1999)
Evidence for quantum critical behavior in the optimally doped cuprate
Bi$_2$Sr$_2$CaCu$_2$O$_{8+\delta}$.
{\it Science} 285:2110-2113.

\bibitem{Sudip1}
Kopp A, Chakravarty S (2005)
Criticality in correlated quantum matter.
{\it Nat Phys} 1:53-56.

\bibitem{Kubo}
Kubo Y, Shimakawa Y, Manako T, Igarashi H (1991)
Transport and magnetic properties of Tl$_2$Ba$_2$CuO$_{6+\delta}$ showing a $\delta$-dependent gradual transition from an 85-K superconductor to a nonsuperconducting metal.
{\it Phys Rev B} 43:7875-7882.

\bibitem{Varma}
Varma CM, Littlewood PB, Schmitt-Rink S, Abrahams E, Ruckenstein AE (1989)
Phenomenology of the normal state of Cu-O high-temperature superconductors.
{\it Phys Rev Lett} 63:1996-1999.

\bibitem{Anderson2006}
Anderson PW (2006)
The `strange metal' is a projected Fermi liquid with edge singularities.
{\it Nat Phys} 2:626-630.

\bibitem{Tallon}
Tallon J, Loram JW (2001)
The doping dependence of $T^*$ - what is the real high-$T_{\rm c}$
phase diagram?
{\it Physica C} 349:53-68.

\bibitem{Hussey}
Hussey NE, Abdel-Jawad M, Carrington A, Mackenzie AP, Balicas L (2003)
A coherent three-dimensional Fermi surface in a
high-transition-temperature superconductor.
{\it Nature} 425:814-817.

\bibitem{Shibauchi2001}
Shibauchi T, Krusin-Elbaum L, Li M, Maley MP, Kes PH (2001)
Closing the pseudogap by Zeeman splitting in
Bi$_2$Sr$_2$CaCu$_2$O$_{8+y}$ at high magnetic fields.
{\it Phys Rev Lett} 86:5763-5766.

\bibitem{Krusin}
Krusin-Elbaum L, Shibauchi T, Mielke CH (2004)
Null orbital frustration at the pseudogap boundary in a layered cuprate superconductor.
{\it Phys Rev Lett} 92:097005.

\bibitem{Basov}
Basov ND, Timusk T (2005)
Electrodynamics of high-$T_{\rm c}$ superconductors.
{\it Rev Mod Phys} 77:721-779.

\bibitem{Abdel}
Abdel-Jawad M {\it et al.} (2006)
Anisotropic scattering and anomalous normal-state transport in a high-temperature superconductor.
{\it Nat Phys} 2:821-825.

\bibitem{Proust}
Proust C, Boaknin E, Hill RW, Taillefer L, Mackenzie AP (2002)
Heat transport in a strongly overdoped cuprate: Fermi liquid and a pure $d$-wave BCS superconductor.
{\it Phys Rev Lett} 89:147003.

\bibitem{Morozov}
Morozov N {\it et al.} (2000)
High-field quasiparticle tunneling in Bi$_2$Sr$_2$CaCu$_2$O$_{8+\delta}$: Negative magnetoresistance in the superconducting state.
{\it Phys Rev Lett} 84:1784-1787.

\bibitem{Shibauchi2003}
Shibauchi T, Krusin-Elbaum L, Blatter G, Mielke CH (2003)
Unconventionally large quantum-dissipative gap regime in overdoped Bi$_2$Sr$_2$CaCu$_2$O$_{8+y}$.
{\it Phys Rev B} 67:064514.

\bibitem{Mackenzie}
Mackenzie AP {\it et al.} (1993)
Resistive upper critical field of Tl$_2$Ba$_2$CuO$_6$ at low temperatures and high magnetic fields.
{\it Phys Rev Lett} 71:1238-1241.

\bibitem{Abrikosov}
Abrikosov AA (2000)
Linear $c$-axis magnetoresistance in underdoped YBa$_2$Cu$_3$O$_{6+\delta}$.
{\it Phys Rev B} 61:5928-5929.

\bibitem{Schofield}
Schofield AJ and Cooper JR (2000)
Quasilinear magnetoresistance in an almost two-dimensional band structure.
{\it Phys Rev B} 62:10779-10784.

\bibitem{Ashcroft}
See, for example, Ashcroft NW and Mermin ND (1976)
{\it Solid State Physics} (Holt, Rinehart and Winston, New York).

\bibitem{Kadowaki}
Tsujii N, Kontani H, Yoshimura K (2005)
Universality in heavy Fermion systems with general degeneracy.
{\it Phys Rev Lett} 94:057201.

\bibitem{Maeno}
Maeno Y {\it et al.} (1997)
Two-dimensional Fermi liquid behavior of the superconducting Sr$_2$RuO$_4$.
{\it J Phys Soc Jpn} 66:1405-1408.

\bibitem{Paglione}
Paglione J {\it et al.} (2003)
Field-induced quantum critical point in CeCoIn$_5$.
{\it Phys Rev Lett} 91:246405.

\bibitem{Bianchi}
Bianchi A, Movshovich R, Vekhter I, Pagliuso PG,
Sarrao JL (2003)
Avoided antiferromagnetic order and quantum critical point in CeCoIn$_5$.
{\it Phys Rev Lett} 91:257001.

\bibitem{Movshovich}
Ronning F {\it et al.} (2006)
Pressure study of quantum criticality in CeCoIn$_5$.
{\it Phys Rev B} 73:064519.

\bibitem{Moriya}
Moriya T, Takimoto T (1995)
Anomalous properties around magnetic instability in heavy electron systems.
{\it J Phys Soc Jpn} 64:960-969.

\bibitem{Tanatar}
Tanatar MA, Paglione J, Petrovic C, Taillefer L (2007)
Anisotropic violation of the Wiedemann-Franz law at a quantum critical point.
{\it Science} 316:1320-1322.

\bibitem{Pham}
Pham LD, Park T, Maquilon S, Thompson JD, Fisk Z (2006)
Reversible tuning of the heavy-Fermion ground state in CeCoIn$_5$.
{\it Phys Rev Lett} 97:056404.

\bibitem{Lake}
Lake B {\it et al.} (2001)
Spins in the Vortices of a High-Temperature Superconductor.
{\it Science} 291:1759-1762.

\bibitem{Wilson}
Wilson SD {\it et al.} (2007)
Quantum spin correlations through the superconducting-to-normal phase transition in electron-doped superconducting Pr$_{0.88}$LaCe$_{0.12}$CuO$_{4-\delta}$.
{\it Proc Natl Acad Sci USA} 104:15259-15263.

\bibitem{vortexNMR}
Kakuyanagi K, Kumagai K, Matsuda Y, Hasegawa M (2003)
Antiferromagnetic vortex core in Tl$_2$Ba$_2$CuO$_{6+\delta}$ studied by nuclear magnetic resonance.
{\it Phys Rev Lett} 90:197003. 

\bibitem{Scalapino}
Scalapino DJ (1995)
The case for $d_{x^2-y^2}$ pairing in the cuprate superconductors.
{\it Phys Rep} 250:330-365.

\bibitem{Wakimoto}
Wakimoto S {\it et al.} (2004)
Direct relation between the low-energy spin excitations and
superconductivity of overdoped high-$T_{\rm c}$ Superconductors.
{\it Phys Rev Lett} 92:217004.

\bibitem{Kopp}
Kopp A, Ghosal A, Chakravarty S (2007)
Competing ferromagnetism in high temperature copper oxide superconductors.
{\it Proc Natl Acad Sci USA} 104:6123-6127.

\bibitem{Hasegawa}
Hasegawa M, Takei H, Izawa K, Matsuda Y (2001)
Crystal growth techniques for Tl-based cuprate superconductors.
{\it J Cryst Growth} 229:401-404.

\bibitem{Kawakami}
Kawakami T, Shibauchi T, Terao Y, Suzuki M, Krusin-Elbaum L (2005)
Evidence for universal signatures of Zeeman-splitting-limited pseudogaps in superconducting electron- and hole-doped cuprates.
{\it Phys Rev Lett} 95:017001.


\end{thebibliography}
\end{document}